\documentclass[a4paper, 12pt,reqno]{amsart}
\usepackage[top=1in, bottom=1in, left=1in, right=1in]{geometry}
\usepackage{floatrow}
\newfloatcommand{capbtabbox}{table}[][\FBwidth]

\reversemarginpar
\usepackage{todonotes}
\usepackage{xcolor}\definecolor{tum_blue}{RGB}{0, 115, 207}\colorlet{col_section }{tum_blue}

\usepackage{tikz}
\tikzset{
  treenode/.style = {shape=rectangle, rounded corners,
                     draw, align=center,
                     top color=white, bottom color=blue!20},
  root/.style     = {treenode, font=\normalsize, bottom color=blue!30},
  decision/.style      = {treenode, font=\normalsize, bottom color=red!30},
  env/.style      = {treenode, font=\normalsize},
  dummy/.style    = {circle,draw}
}
\usetikzlibrary{shapes,arrows}
\usepackage{
	booktabs
	,lmodern
	,multirow
	,subfigure
	,graphicx
	,amssymb
	,amsfonts
	,amsmath
	,amsthm
	,amssymb
	,stmaryrd
	,color
	,array
	,enumerate
    ,microtype
    ,geometry
    ,graphics
	,threeparttable
	,longtable
	,rotating
	,lscape
	,tabularx
	,epsfig
    ,epstopdf
	,setspace
}
\usepackage{floatrow}

\usepackage{chngcntr}
\counterwithin{table}{section}
\counterwithin{figure}{section}
\usepackage[tableposition=top]{caption}
\newcolumntype{Y}{>{\centering\arraybackslash}X}

\usepackage[multiple]{footmisc}

\usepackage{hyperref}
\usepackage{breakurl}
\hypersetup{
	pdfstartview = FitH,
	pdfauthor = {...},
	pdftitle = {...},
	pdfkeywords = {...; ...; ...; ...},
	linktocpage=true,
	colorlinks=true,      
    linkcolor=blue,       
    citecolor=blue,       
    filecolor=blue,       
    urlcolor=blue         
}

\usepackage{xurl}
\usepackage{chapterbib}
\usepackage{natbib}
\newcommand\blfootnote[1]{%
  \begingroup
  \renewcommand\thefootnote{}\footnote{#1}%
  \addtocounter{footnote}{-1}%
  \endgroup
}
 \usepackage{amsaddr}
\newcommand\independent{\protect\mathpalette{\protect\independenT}{\perp}}
\def\independenT#1#2{\mathrel{\rlap{$#1#2$}\mkern2mu{#1#2}}}

\newtheorem{proposition}{Proposition}
\newtheorem{definition}{Definition}

\begin{document}

\title[test EVT1]{Is the distribution of resolvable uncertainty Type I extreme value? A Test for Random Coefficient Models using Choice Probabilities}
\author{Romuald M\'{e}ango\textsuperscript{1}}
\address{\textsuperscript{1} University of Oxford and CESifo}
\noindent \blfootnote{\scriptsize{Correspondence address: Romuald M\'eango, Department of Economics, University of Oxford, Manor Road Building, Manor Road, Oxford, OX1 3UQ, United Kingdom; email: romuald.meango@economics.ox.ac.uk.}}
\date{This version: \today} 

\maketitle

\begin{abstract}
 Stated choice probabilities are increasingly used in conjunction with the random-coefficient model (RCM) to describe individual preferences. They allow survey respondents to express uncertainty about the future or the incompleteness of a hypothetical scenario: the resolvable uncertainty. Parametric assumptions such as a Type I extreme value (EV1) distribution are almost always imposed on this uncertainty to identify and estimate the associated RCM. This paper proposes the first test for these parametric assumptions, based on a nonparametric identification result for the population distribution of the interquantile range of the resolvable uncertainty. In all four empirical applications considered, the test finds strong evidence against the EV1 assumption.

\end{abstract}
\onehalfspacing

\clearpage

\section{Introduction}

McFadden's Additive Random Utility Models have been the workhorse models to analyse consumers/agents' discrete choices and preferences. They are often used under the assumption that there is an unobserved taste shock with Type I extreme value distribution (EV1), although identification and estimation is possible under weaker assumptions \citep[see, for example,][]{manski1988, matzkin1991a,matzkin1992, matzkin1993}. The EV1 assumption results in the tractable mixed-logit model, when the utility has random coefficients \citep{mcfadden2000}. This approach has been used to study revealed preference, where consumers' choices are observed, as well as stated preferences, where survey respondents report their intended choice in a given situation. See the introductory discussion in \cite{blass2010} and a useful review of the methods in \cite{benakiva2019}.


One of the most promising developments of stated preference analyses is the use of probabilistic stated choices, that is, on a scale from 0 to 100 rather than binary choices. This approach has the advantage of allowing respondents to express uncertainty about their intended choice \citep{juster1966, manski1999}. In an influential contribution, \cite{blass2010} operationalise this idea to estimate consumers' preference for electricity reliability in a random utility model. They analyse consumer choice probabilities between $J$ hypothetical alternatives by defining for consumer $i$'s utility for good $j$ a standard random-coefficient model (RCM):
\begin{equation}\label{eq:definition_Uij}
    U_{ij} =  \dfrac{1}{\sigma_i}\left({\gamma_i Y_{ij} + X_{ij}\beta_i + \epsilon_{ij}} \right), J =1, \ldots, J, \text{ with } \sigma_i, \gamma_i >0
\end{equation}
where $X_{ij}$ is a specified function of observed alternative characteristics and personal attributes, and $\beta_i$ are individual specific preference parameters. As is customary in RCMs that translate preference parameters into willingness-to-pay, respondent income (or log-income) $Y_{ij}$ is part of the utility shifters. In a standard choice experiment, $\epsilon_{ij}$ is a utility component that is \textit{observed} by the decision maker but not by the researcher. The authors argue that because choice experiments are often incomplete, the agent may not know $\epsilon_{ij}$ at the time of elicitation but would observe it when s/he makes a decision. This is the resolvable uncertainty. Eliciting choice probabilities allows respondents to express this uncertainty.

In an approach similar to the mixed-logit model, \cite{blass2010} assume that the resolvable uncertainty is distributed as an EV1 variable with standard deviation $\sigma_i$, giving rise to a convenient estimating equation:
\begin{equation}\label{eq:Lad}
    \log \left({\dfrac{P_{ij}}{P_{ik}}}\right) = (Y_{ij} - Y_{ik})\dfrac{\gamma_i}{\sigma_i}+(X_{ij} - X_{ik})\dfrac{\beta_i}{\sigma_i},
\end{equation}
where $P_{ij}$ is $i$'s stated probability of choosing option $j$. The above is used to estimate the mean preference parameters, for example $b = \mathbb{E}(\beta_i/\sigma_i)$, by least-absolute-deviation (LAD). This parameter describes preferences and is often translated into a willingness-to-pay (WTP) for the choices attributes. \cite{wiswall2018} develop this idea further by noting that stated choice experiments generate pseudo-panel data by eliciting choices in several scenarios for the same individual. Thus, an estimate of $(\beta_i/\sigma_i, \gamma_i/\sigma_i)$ can be obtained for each individual and the researcher can estimate a population distribution of WTP (a function of $\beta_i/\gamma_i$).

The assumption of an EV1 distribution is now standard in the literature on stated preferences and is used with hypothetical and nonhypothetical scenarios \citep[Recent examples include][]{arcidiacono2020, aucejo2023, boneva2022, kocsar2022, wiswall2021}. 
Although it is key for the validity of the proposed estimation approaches, it has not been subject to much scrutiny.

This research proposes a test for the EV1 assumption within the RCM framework. The test makes use of a definition of ex ante returns, say $S_{ijk}$, the minimum pecuniary compensation needed by the agent to choose $k$ rather than option $j$. Under the RCM framework and the assumption of an EV1 distribution of the resolvable uncertainty, the conditional distribution of $S_{ijk}$ is a logistic distribution. This imposes restrictions on the quantiles and the interquantile range (IQR) of $S_{ijk}$. In fact, suppose that $S_{ijk}$ has mean $\mu_{Si}$ and standard deviation $\sigma_{Si}$, then the quantiles of $S_{ijk}$ are defined by $\mu_{Si} + \sigma_{Si} \log\left({\dfrac{\tau}{1-\tau}}\right), \tau \in (0,1)$, and the interquantile range, $IQR_i(\tau,0.5)$, the distance of a $\tau$-quantile from the median is $\sigma_i \left|{\log\left({\dfrac{\tau}{1-\tau}}\right)}\right|$. Thus, when we consider the population of respondents, the distribution of 
\begin{equation}\label{eq:def_IQRell}
    IQR_i(\tau,0.5)/\ell(\tau) \text{ with } \ell(\tau):=\left|{\log\left({\dfrac{\tau}{1-\tau}}\right)}\right|
\end{equation} reflects the distribution of $\sigma_{Si}$ in the population and is invariant with $\tau$. This is our main testable implication.\footnote{For a normally distributed resolvable uncertainty, the equivalent correction to $\ell(\tau)$ should be $\left|{\Phi^{-1}(\tau)}\right|$.}$^{,}$\footnote{
The usual approaches to testing the EV1 assumption with revealed preference data include (i) likelihood ratio tests comparing the mixed-logit model to a more general model (which relaxes the EV1 assumption), and (ii) testing for Independence of Irrelevant Alternatives (IIA). For example, the Hausman-McFadden \citep{hausman1984} or the Small-Hsiao test \cite{small1985} compare estimates from the entire set of choices with those from a subset of choices. Unlike these approaches, we do not need a competing parametric model and we require only two choice alternatives. The proposed approach aligns with general practices in model validation that analyse residuals to assess the fit of assumed distributions.}

The paper also investigates the weaker assumption that the difference $\epsilon_{ij} - \epsilon_{ik}$ is symmetrically distributed.\footnote{\cite{blass2010} propose as an alternative assumption that the resolvable uncertainty has median zero and the distribution of $\beta_i$ is symmetric. This is not necessarily nested in our assumption.} The latter assumption imposes $IQR_i(\tau,0.5) = IQR_i(1-\tau,0.5)$ for all $\tau \in (0,1)$. This is our second testable implication. 
To exploit these implications, we need (i) a (nonparametric) identification result of the population distribution of IQR and (ii) a test of equality between the population distributions of IQR.

Section \ref{sec:model} presents the theoretical framework and builds on \cite{meango2024} to provide a nonparametric characterisation for the distribution of (\ref{eq:def_IQRell}) in the RCM model.

Section \ref{sec:test} introduces the test strategy for the EV1 and the Symmetry assumptions. It is based on the test for moment equalities of \cite{andrews2010}, which cumulates the distance between the distributions of interest and checks if this distance is significantly larger than 0. The critical value for the test is obtained by simulation. It is important to note that the test does \textit{not} require panel data.

Section \ref{sec:empirics} provides four empirical applications: The first uses a stated choice experiment on preferences for job attributes of New York University (NYU) students \citep[][WZ2018 hereafter]{wiswall2018}. The second uses the stated choice experiment on preferences for job attributes of students in elite universities in C\^ote d'Ivoire \citep[][MG2025]{meango2024}. The third uses an experiment introduced in two waves
of the New York Fed's Survey of Consumer Expectations to investigate how migration and location choice decisions depend on a set
of location characteristics \citep{kocsar2022}. Finally, the fourth uses the stated choice experiment in \cite{aucejo2023} on the intended likelihood that Arizona State University (ASU) students enrol in higher education at different
costs and possible states of the world. The hypothetical states are related to the COVID-19 pandemic and vary in terms of class formats and restrictions to campus social life. In all four applications, we reject the assumption of an EV1 uncertainty. In the choice experiment for NYU students, job options are symmetric except for the presented attributes. In this case, we do not reject the weaker assumption of symmetry. This section ends by showing that the WTP estimates in the WZ2018 application derived under the rejected EV1 assumption differ from the estimate of the nonparametric identification strategy introduced in MG2025 with a magnitude that is economically relevant.

Section \ref{sec:conclusion} concludes. 

\section{Random coefficient model of probabilistic choice and identification of the IQR}\label{sec:model}
Suppose a binary choice set $\{0,1\}$.\footnote{The framework applies to polychotomous choices as well, with a minor change of notation (MG2025). Note that under the EV1 assumption, discrete choices with more than two choice options can be construed as a set of pairwise choices. The elicitation of choice probabilities rather than discrete choice gives information on each pair.} The utility of the agent is described by Equation (\ref{eq:definition_Uij}). Define $S_i$ as the minimum pecuniary transfer that guarantees that $i$ prefers option 0. With the notation in introduction, it is easy to see that:
\begin{equation}
    S_i:= S(w, \beta_i,\gamma_i,\nu_i) = (y_1 - y_0) + (x_1 - x_0) \beta_{i}/\gamma_i + (\epsilon_1 -  \epsilon_0)/\gamma_i
\end{equation}
where $w=(y_1,y_0,x_1,x_0)$ are counterfactual values of $Y_1,Y_0,X_1,X_0$ and $\nu = (\epsilon_1 - \epsilon_0)/\gamma$.
Note that under the assumption that the distribution of $\epsilon_{ij}$ is EV1, for any $w$, the conditional distribution of $S_i$ given $(\gamma_i,\beta_i, \sigma_i)$ is logistic with standard deviation $\sigma_{Si}:= \sigma_i/\gamma_i$. As discussed in the Introduction, this implies that $IQR_S(\tau,0.5) = \sigma_{Si} \ell(\tau)$. This is a useful restriction. Although each individual in the population may have a different resolvable uncertainty in the sense that $\sigma_i$ and $\gamma_i$ are all individual specific, under the EV1 assumption, the population distribution of $IQR_S(\tau,0.5)/\ell(\tau)$ \textit{must} remain invariant.

To exploit this restriction, it is necessary to characterise the distribution of $S_i$ even without the EV1 assumption. We use the framework of MG2025, who show that this is possible using choice probabilities. More formally, during a survey experiment, $i$ is presented with a scenario characterised by $W = (Y_{i1},Y_{i0},X_{i1},X_{i0})$ and is asked to state their chance of choosing option 1 over option 0, say $P_{i}$. Presented with scenario $W_i$, respondent $i$ states:
\begin{equation}
    \label{eq:definition_m}
    P_i = m(W_i,\eta_i) \text { where } m(w,\eta_i):= \Pr(S \left({w,\eta_i,\nu_i}\right) \ge 0 \vert \eta_i) = F_{\nu \vert \eta}(S \left({w,\eta_i,\nu_i}\right) \ge 0 \vert \eta_i).
\end{equation}
$\eta_i$ encompasses $i$'s preferences parameter $(\beta_i,\gamma_i, \sigma_i)$, private information, or any other unobserved characteristic that influences the resolvable uncertainty. The dimension of the random vector $\eta$ is unrestricted. The mapping $w \mapsto m(w,\eta)$ defines the \textit{stated demand function} for an individual with characteristic $\eta$. In line with the literature, the stated choice experiment is construed as a \textit{ceteris paribus} experiment. Within it, the respondents are asked to report their stated choice as if $W_{i}$ was determined exogenously. They do not infer new unspecified attributes as the specified attributes change. In practice, to ensure that it holds, the survey design includes explicit or implicit instructions for the respondents.

To see the importance of this assumption, it is instructive to compare equation (\ref{eq:definition_m}) with a more general definition of elicited preferences:
    \begin{equation}\label{eq:definition_non_cpa}
        P_{i} = \Pr \left({S \left({W_i,\eta_i,\nu_i}\right) \ge 0 \Bigr| W_{i},\eta_i }\right).
    \end{equation}
In equation (\ref{eq:definition_non_cpa}), the perceived distribution of $\eta^*$ changes with $W_{i}$ and the respondent uses it to infer a distribution of resolvable uncertainty. This is sometimes called the \textit{fill-in problem} \citep[see, for example,][]{hudomiet2018}. It makes it impossible to distinguish between the effect of the choice attributes on the preferences and the beliefs. The assumption of a ceteris paribus experiment is key because it ensures that the analyst can take advantage of the variation in the choice attributes to understand the preferences. This framework includes the EV1 assumption where the variance is determined by an individual specific parameter $\sigma_i$ as a special case, but imposes a weaker structure.

Our interest lies in the individual-specific distribution $\Pr\left({S \left({w,\eta_i,\nu_i}\right) \le s \vert \eta_i}\right)$, or more specifically in its quantiles, and how these are distributed in the population. Theorem 1 of MG2025 provides a characterisation of the population distribution of quantiles of $S_i$ using the conditional distribution of choice probabilities $P_i$. Proposition \ref{prop:characterisation} builds on this result to provide a characterisation of the distribution of IQR that serves in the test.
\begin{definition}
    Let $P_i$ be defined by Equation (\ref{eq:definition_m}) and let $F_{\nu|\eta}$ be a continuous distribution. Let $F_{\tilde{W}}$ be the cumulative distribution function of the variable $\tilde{W}$, which is of interest to the analyst. For example, the analyst may be interested in setting all attributes to be same except one, to understand the WTP for this attribute.

    We denote with $Q_{P|X}$ the conditional (on X) quantile of the variable $P$.
    Define:
    $F_{S,i}(s;\tilde{W}) := \Pr(S(\tilde{W},\beta_i,\gamma_i,\nu_i) \le s | \eta_i)$ and     
    $Q_{S,i}(\tau;\tilde{X}):=\inf \{s: F_{S,i}(s;\tilde{W}) \ge \tau\}$, respectively the individual-specific distribution of returns and its quantile.    
    Define also $F_{Q}(s;\tau, F_{\tilde{W}}):= \Pr\left({Q_{S,i}(\tau;\tilde{W}) \le s }\right)$ the distribution of quantiles in the population.
    
    Furthermore, for any $0<\tau_1< \tau_2<1$, the interquantile range for an individual with scenario $\tilde{W}$ and unobserved characteristic $\eta$ is defined by:
\begin{equation}\label{eq:definition_IQR}
IQR(\tau_1,\tau_2;\tilde{W},\eta)= Q_{S}(\tau_2;\tilde{W},\eta) - Q_{S}(\tau_1;\tilde{W},\eta)
\end{equation}
    Finally, denote $t(s,w) = (y_1-s,y_0,x_1,x_0)$.
\end{definition}
\begin{proposition}\label{prop:characterisation}
    If $W \independent \eta$, the following holds:\vskip4pt
For any real value $s$ such that $t(s,w) \in \mathcal{W}$ and $\tau \in [0,1]$,
\begin{eqnarray}\label{eq:ident_FQ}
\nonumber F_{Q}(s;\tau, F_{\tilde{W}}) =  \int_{\mathcal{W}} \int_0^1 1\left\{{Q_{P \vert W}(a \vert t(s,w)) \le 1- \tau}\right\}\; da \;dF_{\tilde{W}}(w)
\end{eqnarray}
Let:
\begin{equation}\label{eq:QTR_IQR}
   A^{\tau_2 - \tau_1}(w,a) = \int_{\mathcal{S}}  1 \left\{{1-\tau_2 \le Q_{P \vert W}(a|t(s,w)) \le 1-\tau_1}\right\}ds 
\end{equation}
The population distributions of IQR is identified by:
\begin{equation}
\Pr\left({IQR(\tau_1,\tau_2;\tilde{W},\eta) \le y }\right)
=  \int_{\mathcal{W}} \int_{0}^1   1 \left\{{A^{\tau_2 - \tau_1}(w,a) \le y }\right\}  da\; dF_{\tilde{W}}(w). \label{eq:ident_IQR}
\end{equation}
\end{proposition}
The first part of Proposition \ref{prop:characterisation} is proved in MG2025 and shows that the conditional distribution of $P$ given $W$ can be used to retrieve the population distributions of the quantiles of $S$ and the IQR of $S$, for all values of their support such that $t(s,w) \in \mathcal{W}$, the support of $W$. The main requirement is that $W$ is independent of $\eta$ (MG2025, Assumption 1) and that the experiment is ceteris paribus (MG2025, Assumption 2). The former condition holds for stated choice experiments where the scenarios are chosen exogenously. The second condition holds when instructing respondents to consider changes in the observed attributes only. The continuity and monotonicity requirements of MG2025 are satisfied in the RCM framework.

The intuition for this result is the following: The analyst is interested in the distribution  $\Pr(S(w,\eta_i,\nu_i) \le s \vert \eta)$, the distribution of ex ante returns. The respondent provides information about: $\Pr(S(W_{i},\eta_i,\nu_i) \ge 0 \vert \eta_i):= m(W_{i},\eta_i)$, the \textit{stated demand}. Because the choice attributes $W_{i}$ vary exogenously, the analyst can use the variations in $W_{i}$ to learn about the distribution of interest. In particular:
\begin{eqnarray}
    \Pr(S(y_1,y_0,x,\eta,\nu) \le s \vert \eta) &=& \Pr(S(y_1-s,y_0,x,\eta,\nu) \le 0 \vert \eta), \text{ by linearity of } S,\\
    &=& \Pr(S(Y_1,Y_0,X,\eta,\nu) \le 0 \vert Y_1 = y_1-s,Y_0 = y_0,W,\eta)\\
    && \text{ by independence,}\\
    &=& 1-m(t(s,W),\eta), \text{ by definition of }m.
\end{eqnarray} The stated demand function $m$ is not directly identified, but its quantile treatment response function corresponds to $Q_{P|W}$ due to the independence between $W$ and $\eta$. The population distribution of equation (\ref{eq:ident_FQ}) is obtained by integrating over all quantiles. As discussed in MG2025, this result holds even without the linearity structure of the RCM.

The following provides a proof of the second part of Proposition \ref{prop:characterisation}. We note that: $ Q_{P|W}(a|w) = Q_{m(w,\eta)}(a)$, the quantile treatment response (QTR) function. For any $a \in (0,1)$ and any $w \in \mathcal{W}$
    \begin{eqnarray*}
Q_{m(w,\eta) }(a) &=& Q_{m(W,\eta) \vert W}(a \vert w) \text{ by the independence between }W \text{ and } \eta\\
        &=& Q_{P \vert W}(a \vert w)
    \end{eqnarray*}
The QTR is very convenient because it translates a possibly infinite dimensional problem (the dimension of $\eta$) to a uni-dimensional problem. Indeed, for any respondent $i$, there exists $\alpha_i$, a realisation of a uniformly distributed random variable such that $P_{i} = m(W_{i},\eta_i) =q(W_{i},\alpha_i)$, where $\alpha_i = F_{P \vert W}(P_{i} \vert W_{i})$.
\vskip4pt
We have established above that $\Pr(S(w,\eta,\nu) \le s \vert \eta) = 1-m(t(s,W),\eta).$ This represents a cumulative distribution function and one can derive the conditional quantiles, say  $Q_{S}(\tau;x, \eta)$, as in \cite{chernozhukov2020} \citep[cf. also][pp. 113-114]{karr1993}. Let $\mathcal{S}$ denote the support of $S$. 
\begin{eqnarray}
Q_{S}(\tau;w,\eta) &=& \int_{\mathcal{S}} \left\{{\left[{ 1- m(t(s,w),\eta)}\right] \le \tau}\right\} - 1\{s\le 0\} ds
\label{eq:average_return}
\end{eqnarray}
$Q_{S}(\tau;w,\eta)$ can be seen as random objects, which is a strictly increasing functional of $m(w,\eta)$. 
Note also that by equations (\ref{eq:definition_IQR}) and (\ref{eq:average_return}), the $IQR$ is such that:
\[IQR(w,\eta;\tau_1,\tau_2)= \int_{\mathcal{S}}  1 \left\{{1-\tau_2 \le m(t(s,w),\eta) \le 1-\tau_1}\right\}ds\]
which is also a strictly increasing functional of $m$. 
Recall that: \[
\begin{array}{l}
A^{\tau_2 - \tau_1}(w,a) = \int_{\mathcal{S}}  1 \left\{{1-\tau_2 \le Q_{P|X}(a|t(s,w)) \le 1-\tau_1}\right\}ds
\end{array}
\]
Thus, $A^{\tau_2 - \tau_1}(w,a)$ that replaces $Q_{P|X}(a|t(s,w))$ for $ m(t(s,w),\eta)$ represents the QTR for the $IQR(w,\eta;\tau_1,\tau_2)$.

Therefore, the population distribution of IQR can be rewritten can be rewritten:
\begin{eqnarray*}
\Pr\left({IQR(\tau_1,\tau_2;W,\eta) \le y }\right)
=  \int_{\mathcal{W}} \int_{0}^1   1 \left\{{A^{\tau_2 - \tau_1}(w,a) \le y }\right\}  da\; dF_{\tilde{W}}(w).
\end{eqnarray*}
This completes the proof.

With the nonparametric characterisation of the population distribution of IQR in hand, we can now test whether it is invariant after applying the normalisation $\ell(\tau)$ or satisfies symmetry. The restrictions imposed by the EV1 and the symmetry assumptions amount to a set of unconditional moment equalities. The test can be performed using an existing procedure such as \cite{andrews2010}.

\section{Test Procedure}\label{sec:test}
Let $G_{\tau}(y) = \Pr(IQR(\tau;0.5) \le y \ell(\tau))$. Consider a collection of $ \mathcal{T} = \{\tau_1, \ldots, \tau_K\}$.\footnote{In practice, because answers to choice probabilities questions tend to be rounded to multiples of 5 or 10, we do not need to consider a continuum of $\tau$.} The null hypothesis for the EV1 assumption is:
\begin{equation}\label{eq:null_TIEV}
    H_0: G_{\tau}(\cdot) = G_{\tau'}(\cdot), \text{ for any } \tau,\tau' \in \mathcal{T}
\end{equation}
The null hypothesis for the symmetry assumption is:
\begin{equation}
    H_0: G_{\tau}(\cdot) = G_{1-\tau}(\cdot), \text{ for any } \tau \in \mathcal{T}.
\end{equation}
The logic of the test procedure is the same for both tests, so we discuss the test for the first hypothesis only. Define for a finite collection $\mathcal{Y}$ in $\mathbb{R}^+$ the long vector $M$ obtained by collecting all differences $G_\tau(y) -G_\tau'(y)$ for $\tau \neq \tau', \tau,\tau' \in \mathcal{T}$ and $y \in \mathcal{Y}$. Let $\dim({M})$ denote its dimension. Equation (\ref{eq:null_TIEV}) implies that: 
\begin{equation}\label{eq:test_statistic}
    S(M,\Omega) := (M'\Omega M)^{1/2} = 0  
\end{equation}
for any conformable, positive, semi-definite matrix $\Omega$.

The proposed test statistic uses the empirical version of $G_{\tau}(\cdot)$, $\widehat G_{\tau,n}(y)$, to construct $\widehat{M}_n$, the empirical counterpart of $M$. For $\Omega$, it uses the inverse of the variance-covariance matrix of the random vector $M$, a matrix of size $\dim(M)\times\dim(M)$. The empirical counterpart $\widehat{\Omega}_n$ is constructed using a bootstrap sample. Finally, the critical value is obtained using the plug-in method in Section 7 of \cite{andrews2010}. Let $Z\sim N(0,\Omega)$ be a random vector of the same dimension as $M$. The reader should think of $Z$ as a replication of $M$ under $H_0$. The critical value $c(\Omega, 1-\alpha)$ is obtained as the $(1-\alpha)$- quantile of the distribution of $S(Z,\Omega)$. It is calculated by simulating a large number of random variables $Z_l \sim N(0,\widehat{\Omega}_n)$ and taking the $(1-\alpha)$-quantile of the collection of $\{S(Z_l,\widehat{\Omega}_n)\}_{l=1}^L$. Reject the null hypothesis at level $\alpha$ if $S(\widehat{M}_n,\widehat{\Omega}_n)$ exceeds the calculated critical value. The detailed procedure follows.

Let $\{P_i,W_i\}_{i=1,\ldots,N}$ be the i.i.d. sample observed by the analyst. Denote by $\{P_i^b,W_i^b\}_{i=1,\ldots,N}$ a bootstrap sample obtained by resampling. The test procedure is as follows:
\vskip4pt
\paragraph{\textit{Step 1.}} Estimate the quantile treatment response of $P$ given $W$,  $Q_{P\vert W}(.|.)$ in the original data $\widehat{Q}_{P\vert W}(.|.)$ and in the bootstrap sample $\widehat{Q}^b_{P\vert W}(.|.)$. In the empirical applications, $\widehat{Q}_{\ell(P)\vert W}(a|w) = r(w)*\hat{\beta}_a$ is estimated by performing several quantile regressions, where $r(x)$ is the vector of differences in choice attributes.\footnote{It is possible to consider the interactions, but we choose to remain as close as possible to the simple linearity structure.} To reduce computation time and mitigate the effect of rounding, the regression is estimated on quantiles 0.01, 0.5, 0.15, \ldots, 0.95, 0.99. For the remaining quantiles, the regressions coefficient are interpolated, using linear interpolation with the Matlab routine `griddedInterpolant'.
\vskip4pt
\paragraph{\textit{Step 2.}} For each $\tau \in \mathcal{T}, \tau >0.5, y \in \mathcal{Y}$ estimate the empirical counterpart of the population distribution of $IQR$ defined by Equation (\ref{eq:ident_IQR}).
\begin{eqnarray}
    \widehat{G}_{\tau,n}(y) &=& \dfrac{1}{N}\sum_i \sum_{k = 1}^{K_a} \delta_a 1 \left\{{ \widehat{A}_n^{\tau -0.5}(W_i,a_k) \le y \ell(\tau) }\right\} \label{eq:Gtau} \\
    \widehat{A}_n^{\tau -0.5}(w,a) &=& \sum_s \delta_s 1 \left\{{1-\tau \le \widehat{Q}_{P\vert W}(a|t(s,w)) \le 0.5}\right\} \label{eq:Atau}
\end{eqnarray}
The first sum is on a finite grid of the unit interval with $K_a$ points $\{a_1,\ldots,a_{K_a}\}$ and step width $\delta_a$. The sum over $s$ in the expression of $\widehat{A}^{\tau -0.5}(w,a)$ is on a fine grid of $[0,K_S\times \delta_s]$ where $K_S$ is a large integer and step width $\delta_s>0$. There are similar expressions for $\tau >0.5$.
\vskip4pt
\paragraph{\textit{Step 3.}} Construct the matrix $\widehat{M}_n$, the matrix collecting all differences $\widehat{G}_{\tau,n}(y) -\widehat{G}_{\tau',n}(y)$ for $\tau \neq \tau', \tau,\tau' \in \mathcal{T}$ and $y \in \mathcal{Y}$ for the original and the bootstrap sample. Stacking the bootstrap sample of $\widehat{M}_n^b$ in a matrix $\widehat{\boldsymbol{M}}_n$ of dimension $B \times \dim(M)$, obtain the empirical estimate $\widehat{\Sigma}_n =\widehat{\text{cov}}\left({\widehat{\boldsymbol{M}}_n}\right)$. Finally, estimate $S(\widehat{M}_n,\widehat{\Sigma}_n^{-1})$ of dimension $\dim(M) \times \dim(M)$.
\vskip4pt
\paragraph{\textit{Step 4.}} Simulate $L$ random variables $Z_l \sim N(0,\widehat{\Sigma}_n)$ and compute $S(Z_l,\widehat{\Sigma}_n^{-1})$. Compute the critical value $c\left({\widehat{\Sigma}_n^{-1},1-\alpha}\right)$ as the $(1-\alpha)$-quantile of the simulated collections $\left\{{S\left({Z_l,\widehat{\Sigma}_n^{-1}}\right)}\right\}\}_{l = 1}^L$.
\vskip4pt
\paragraph{\textit{Step 5.}} Reject $H_0$ if 
$S\left({\widehat{M}_n,\widehat{\Sigma}_n^{-1}}\right) > c\left({\widehat{\Sigma}_n^{-1},1-\alpha}\right)$.
\vskip4pt
The results of \cite{andrews2010} implies that the proposed test procedure is valid and has good power properties under usual regularity assumptions. Note that the uniform central limit theorem for the estimator of $G_{\tau}(\cdot)$ is the result of the uniform central limit theorem for the quantile regression estimator and the Hadamard differentiability of all operators involved \cite[see discussions in][]{meango2024}. Note also that as discussed in \cite{blass2010}, rounding is a small concern for quantile regressions if the respondents are consistent in their rounding behaviour. To mitigate the effect of the pattern of rounding to multiples of 10 often encountered in the stated preference literature, I advise computing the quantile regressions at $0.05,0.15,\ldots,0.95$, and interpolating the coefficients.    

\section{Four empirical applications}\label{sec:empirics}

This Section applies the proposed testing procedure to four stated choice experiments. The first uses the stated choice experiment of WZ2018 on the preferences for job attributes of 247 students from New York University.\footnote{The original survey experiment has two blocks. Here, we consider only the first block. The second block uses a different set of attributes (the chance of being fired, the proportion of males, and the percentage of bonuses in pay). The conclusions of the test are the same for the second block.} Within the stated choice experiment, students are presented with three job options. The job options do not have a specific label and are characterised by the wage, the number of hours worked, the potential for future earning growth, and whether there is the option of a part-time job. Each student is presented with eight scenarios, and the survey records the choice probability for each given alternative. The following testing procedure considers pairwise choices, which is valid under the EV1 assumption.  

The second application uses the stated choice experiment of MG2025 on preferences for job attributes of students in elite universities in C\^ote d'Ivoire. Within the stated choice experiment, 587 students are presented with two job options: One in the public sector and one in the private sector. Each job option is characterised by the wage, the number of hours worked, the probability of losing the job, and the chance of moving up in the hierarchy. Each student is presented with five scenarios, and the survey asks the probability of choosing one of the two alternatives given.

The third application is from \cite{kocsar2022} and uses the stated preference approach to understand residential migration and location decision of a nationally representative sample of US residents. It was introduced in two waves from the New York Fed's Survey of Consumer Expectations (SCE, September 2018 and December 2019). The survey collects data on individuals’ probabilities of moving over the next two years by choosing from a set of hypothetical locations, as well as their current location. It experimentally varies the characteristics of the locations such as income prospects, housing costs, proximity to family and friends, local social norms and values, state and local taxes, etc. in order to identify individuals’ preferences for various location attributes. The survey has 1,861 individuals and about 27,800 individual-scenarios. Each scenario has three choice options (home and two migration options). The following testing procedure considers pairwise choices with the choice of staying home being the reference in each pair. This is valid under the EV1 assumption.

The final application is from \cite{aucejo2023} and estimates the utility that Arizona State University (ASU) students derive from in-person instruction and on-campus social activities. In late April 2020, it elicited the intended likelihood of enrolling in higher education for approximately 1,500 students. The 42 hypothetical scenarios are related to the COVID-19 pandemic and vary in terms of class formats (in-person or remote instruction), restrictions to campus social life, the existence of a vaccine, and the prevalence of COVID-19. The \textit{numeraire} in this experiment is the cost of college, which is also varied exogenously.

\begin{figure}
    \centering
  \subfigure[NYU students]  
  {
  \includegraphics[width=0.45\linewidth]{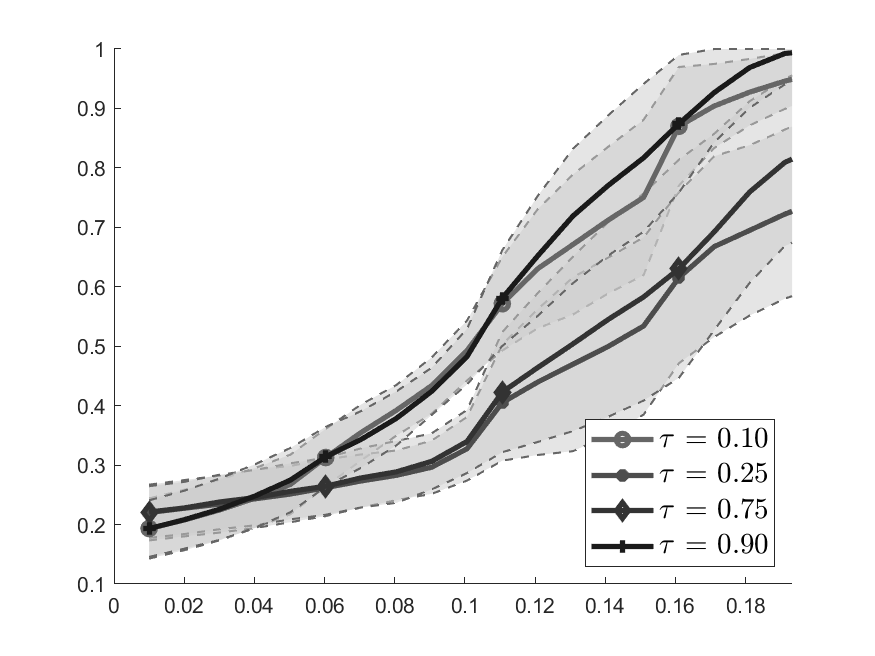}
  }
  \subfigure[Ivorian students]
    {
  \includegraphics[width=0.45\linewidth]{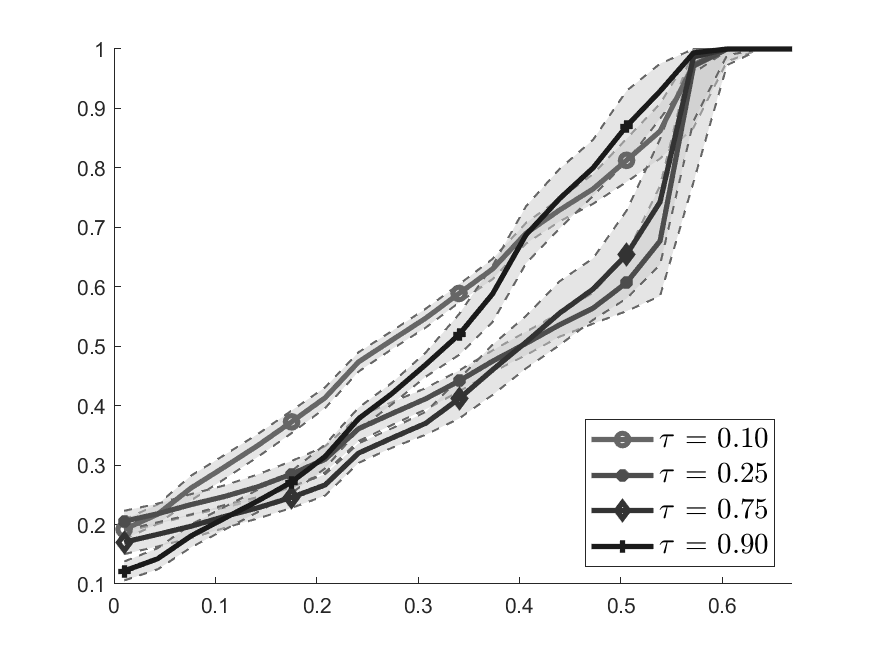}
  }
    \subfigure[SCE panel]
    {
  \includegraphics[width=0.45\linewidth]{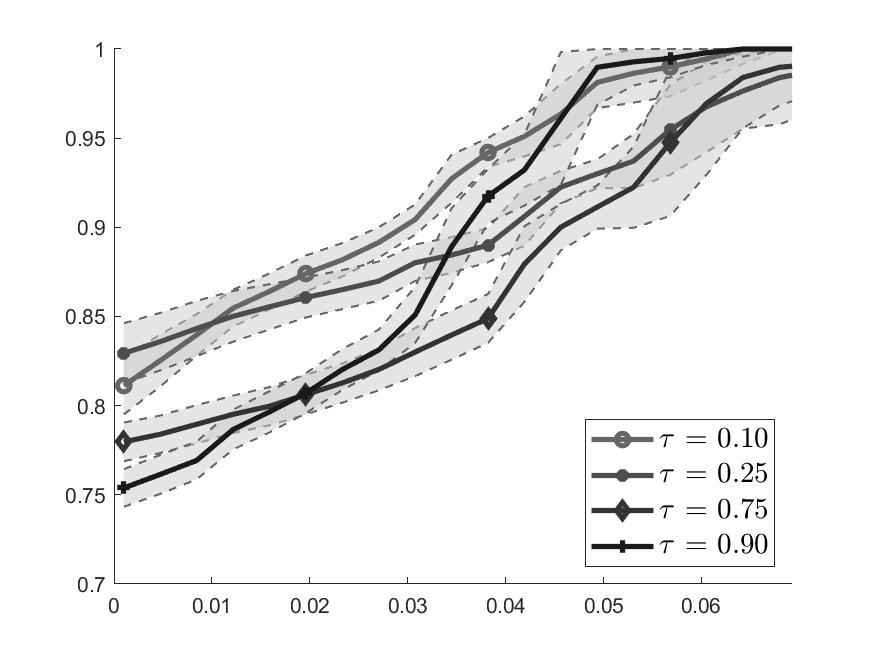}
  }
      \subfigure[ASU Students]
    {
  \includegraphics[width=0.45\linewidth]{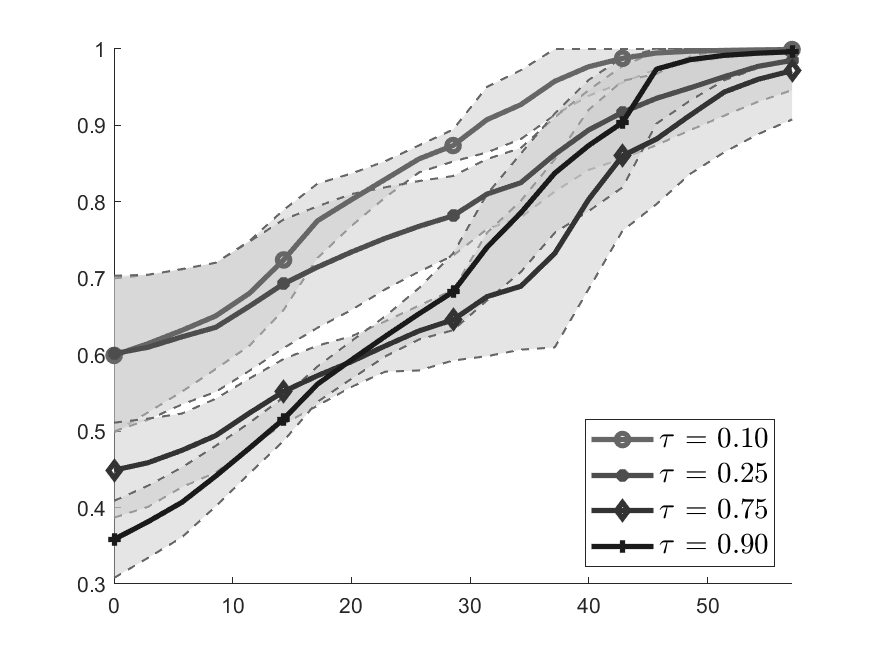}
  }
    \caption{Estimates of the population distributions $G_{\tau}(y)$, for $\tau = 0.1,0.25,0.75,0.90$, with 90\% confidence intervals}
    \label{fig:G_taus}
    \floatfoot{Note: The figures display the estimated cumulative distribution functions $\widehat{G}_{\tau}(\cdot)$ using Equations (\ref{eq:Gtau}) and (\ref{eq:Atau}) and the pointwise confidence region following the procedure of MG2025. The X-axis is the support of $IQR(\tau,0.5)/\ell(\tau)$. The solid lines represent the estimated distributions, and the shaded regions represent the confidence regions. Under the EV1 assumption, the distributions should be the same. The discrepancy may come from sampling error. The test allows to check whether the differences are statistically significant. For the SCE panel, we use a subsample of 10,000 observations, as the implementation on the full saturates the memory on an ordinary computer. }
\end{figure}
Figure \ref{fig:G_taus} displays the estimated cumulative distribution functions $\widehat{G}_{\tau}(y)$ using Equations (\ref{eq:Gtau}) and (\ref{eq:Atau}) and the pointwise confidence region following the procedure of MG2025. The estimation uses 500 bootstrap replications to calculate the standard error. Block bootstrap is used to preserve the panel structure. Regressors $W$ are introduced as the difference between the attributes in the two options. We consider log-earnings and the results are qualitatively similar with income levels. For the ASU sample, we consider a quadratic polynomial of costs. The X-axis is the support of $IQR(\tau,0.5)/\ell(\tau)$. The solid lines represent the estimated distributions and the shaded regions represent the 90\% uniform confidence regions.

We can make four remarks: First, beliefs about the resolvable uncertainty are heterogeneous in the population. This provides empirical evidence against a commonly made assumption `that respondents make the same assumption subjectively' \citep[][p.424]{blass2010} or in the above notations, $\sigma_i = \sigma,$ for all $i$.

Second, the magnitude of the resolvable uncertainty depends on the context. For example, in the NYU sample, it can be large: For the median individual of the $IQR(0.25,0.5)$-distribution, the resolvable uncertainty is between 11 and 15 percent of the earnings in the scenarios. In the SCE sample, close to three quarters of the respondents do not have any uncertainty about the decision to move over the next two years. In the ASU sample, the distributions are compressed in the lower quantiles and have a long right tail.

Third, under the EV1 assumption, all distributions should be the same. This does not appear to be the case at first sight. Of course, the discrepancy may come from sampling error. The test allows to check rigorously whether the differences are statistically significant.

Fourth, symmetry seems to hold in the NYU sample and not in the other samples. In the former, the elements of each pair $(G_{0.10},G_{0.90})$ and $(G_{0.25},G_{0.75})$ are very close. This should be expected because the hypothetical jobs in the NYU sample are not labelled. Thus, the choice options are a priori symmetric, except for the hypothetical attributes presented in the scenario. In contrast, the Ivorian, SCE and ASU samples exhibit a discrepancy, suggesting that respondents perceive differently the amount of resolvable uncertainty in the public and private sectors, for staying home and moving to another location, or for enrolling and not enrolling in college. 

\begin{table}[htbp]
    \centering
    \begin{tabular}{llrrrrl}
    \hline \hline
    &  & \multicolumn{1}{l}{Test} & \multicolumn{3}{l}{Critical values} & \\
Sample  & $H_0$ & \multicolumn{1}{l}{Statistic} & 10\% & 5\% & 1\% & Decision\\
\multicolumn{7}{l}{}\\
\hline
NYU    & EV1    & 51.09 &  35.83 &  36.18  & 36.88 & Reject at 1\% \\
      & Symmetry &   13.64 &  16.49&  16.81 &  17.37 & Do not reject \\
      \multicolumn{7}{l}{}\\ \hline
Ivoirian    & EV1    &   223.19  & 37.64 &   38.01 &  38.68 & Reject at 1\%\\
& Symmetry   & 20.36&  18.11 &  18.44 &  19.06 & Reject at 1\%\\
\multicolumn{7}{l}{}\\  \hline
SCE    & EV1    &   55.91  & 35.86  & 36.19   &36.87& Reject at 1\%\\
& Symmetry   &21.83 & 16.80 &  17.12   &36.87 & Reject at 1\%\\
\multicolumn{7}{l}{}\\
 \hline
ASU    & EV1    &   68.24  & 37.29  & 37.61   &38.33& Reject at 1\%\\
& Symmetry   &18.72  & 17.23 &  17.54   &18.14& Reject at 1\%\\
\multicolumn{7}{l}{}\\
            \hline \hline
    \end{tabular}
    \caption{Test results}
    \begin{tablenotes}
\item {\footnotesize Notes: The Table presents the results of the test procedure for both sample: The test statistic, the critical values at conventional levels, and the test decision. The estimated cumulative distribution functions $\widehat{G}_{\tau,n}(\cdot)$ use Equations (\ref{eq:Gtau}) and (\ref{eq:Atau}). The test statistic uses the empirical counterpart of Equation (\ref{eq:test_statistic}). The simulated critical values use $10,000$ draws for the normal distribution. The test uses 500 replications to compute the standard error of $\widehat{\Sigma}_n$.}
\end{tablenotes}
    \label{tab:test_results}
\end{table}
Table \ref{tab:test_results} shows the results of the test procedures for each sample. It presents the test statistic and the associated critical values at conventional levels. The test unequivocally rejects the assumption of EV1 in all four samples at conventional levels. For each sample, the test statistic is significantly higher than the associated critical values. As expected, the symmetry assumption is not rejected for the NYU sample but in the remaining three samples. 

To sum up, we find strong evidence against the assumption of EV1 resolvable uncertainty in all four contexts, and evidence against the assumption of symmetry where we would not expect it to hold. To be clear, the test is for the joint restrictions of the random utility model (including linearity) and the EV1 assumption. The results imply that both cannot hold simultaneously.

Before concluding, it is important to ask whether the failure of the parametric assumption matters empirically. After all, it is mainly intended to approximate the true distribution, which is unobserved. The following result suggests that it does and should make us cautious about how we approach the tasks of identification and estimation.

\begin{figure}
    \centering
    \subfigure[Adding one more hour]{
    \includegraphics[width=0.7\linewidth]{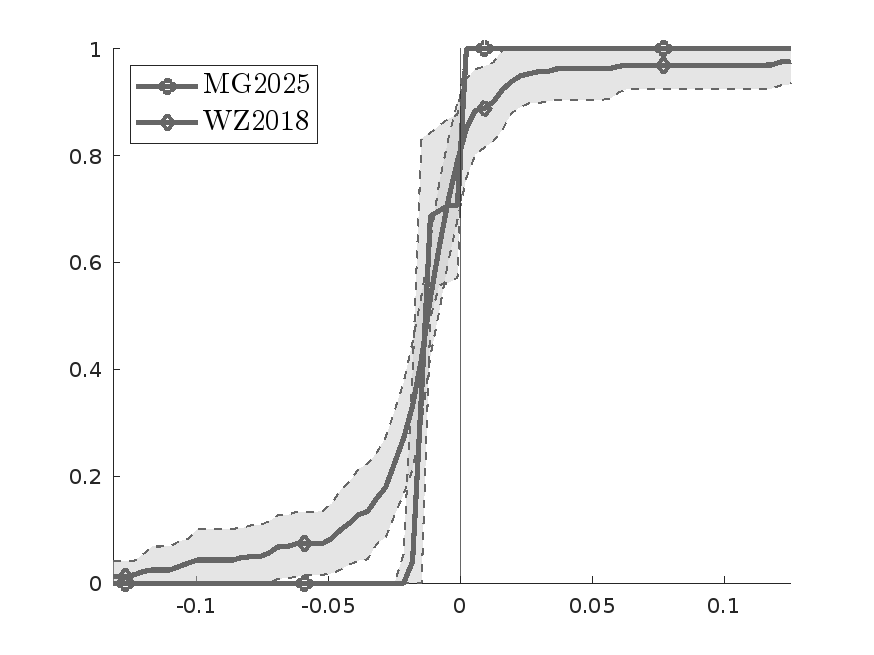}
    }
    \subfigure[Adding the option of working part-time]{
        \includegraphics[width=0.7\linewidth]{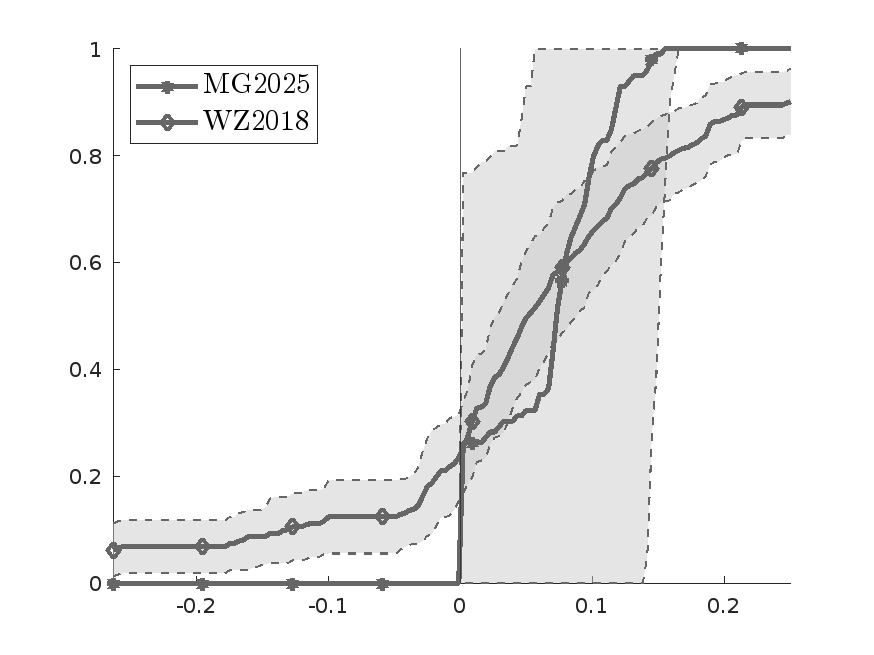}
    }
    \caption{Comparing estimates of women's WTP distributions for changes in attributes.}
    \floatfoot{Notes: The estimated distribution WZ2018 uses the methodology of WZ2018 and \cite{aucejo2023}. It relies on the EV1 assumption to estimate separate LAD regressions for each individual as in Equation (\ref{eq:Lad}). The estimated distribution MG2025 uses Theorem 2 for $qWTP$ and $\tau = 0.5$. Note that under the RCM, rank invariance holds. In panel (a), $F_{\tilde{W}}$ corresponds to an increase of one hour for Option 1, and in panel (b) to the addition of the possibility of working part time. The other differences are muted. The shaded regions represent the 90\% confidence regions. A equality test rejects the null at the significance level 1\%. In panel (a), the test statistic is 41.7, and the critical value at 1\% is 14.0; in panel (b), the test statistic is 23.6, and the critical value at 1\% is 14.0.}
    \label{fig:compare_FS_WZ}
\end{figure}

For the NYU sample, Figure \ref{fig:compare_FS_WZ} shows two estimates of the WTP distribution for two changes in attributes: increasing by one hour the working time and adding the option of working part-time, at the same wage. 
The first estimates use the methodology of WZ2018 that relies on the EV1 assumption to estimate separate regressions as in Equation (\ref{eq:Lad}). Using the least absolute deviation estimator or a fractional response model as in \cite{aucejo2023} yields similar conclusions, and we report the latter results. The second estimates use Theorem 2 for the WTP, $qWTP$, at the median beliefs, $\tau = 0.5$. In panel (a), $F_{\tilde{W}}$ corresponds to an increase of one hour for Option 1, and in panel (b) to the addition of the possibility of working part time, at the same wage. The other differences in attributes are muted. The shaded regions represent the 90\% confidence regions. The semiparametric estimator of MG2025 shows a negative sign for increasing hours of work (between -2\% and 0\% of earnings for most of the population) and a positive sign for adding the flexibility of working part-time (between 0\% and 12\%). In contrast, separate regressions result in distributions that are more dispersed, suggesting opposite effects for part of the population. A test for the equality of the two distributions (similar to our main test) rejects the null at the significance level 1\% in both cases. 

\section{Conclusion}\label{sec:conclusion}
This research proposes to test the restrictions on the resolvable uncertainty in the standard Random Utility Model used with probabilistic stated preference data. It applies to the popular assumption of a Type I extreme-value (EV1) distribution and to the weaker assumption of symmetry, which underlies many other (parametric) distributions. Although the main application is to stated choice experiments, it can be used for nonhypothetical scenarios, provided that the researcher is willing to assume that the observed choice attributes are uncorrelated with the unobserved heterogeneity. The test could also be fruitfully adapted to study other parametric distributions.

In four empirical applications, \cite{wiswall2018}, \cite{meango2024}, \cite{kocsar2022}, and \cite{aucejo2023}, the test finds strong evidence against the EV1 assumption. It does not reject the assumption of symmetry when the choice options are a priori symmetric, except for their hypothetical attributes. These results should encourage researchers to rely on identification strategies obtained under weaker assumptions, such as the maximum score estimation in \cite{blass2010}, or the estimator based on quantile and distribution regressions in \cite{meango2024}.

More generally, the paper raises a broader question for random utility models with revealed preference data. If respondents do not perceive the unobserved heterogeneity to be EV1, we should probably be more cautious about this assumption for the analysis of revealed preference data.

\bibliographystyle{apalike}
\bibliography{ref_expectations}


\end{document}